\documentclass[a4paper,fleqn,usenatbib,referee,epsfig,color]{mnras}

\usepackage[T1]{fontenc}
\usepackage{ae,aecompl}

\usepackage{graphicx}	
\usepackage{amsmath}	
\usepackage{amssymb}	

\newif\ifAMStwofonts
\newcommand{\be}{\begin{equation}}
\newcommand{\ee}{\end{equation}}
\newcommand{\bea}{\begin{eqnarray}}
\newcommand{\eea}{\end{eqnarray}}

\begin{document}
\title[Scaling Relations in Astronomy]{\bf A `Numbers' Approach to Astronomical Correlations
I: Introduction and Application to galaxy Scaling Relations }
\author[R.N. Henriksen ]
{R. N. Henriksen$^1$\thanks{henriksn@astro.queensu.ca} \& J. A. Irwin$^1$\\ 
$^1$Dept. of Physics, Engineering Physics \& Astronomy, Queen's University, Kingston, Ontario, K7L 3N6, Canada\\}
\date{Accepted XXX. Received YYY; in original form ZZZ}
\pubyear{2017}
\label{firstpage}
\pagerange{\pageref{firstpage}--\pageref{lastpage}}
\maketitle

\begin{abstract}
We propose a new systematic method of studying correlations between parameters that describe an astronomical (or any) physical system. We recall that behind  Dimensionless scaling laws  in complex, self-interacting physical objects  lies a rigorous theorem of Dimensional analysis, known widely as the Buckingham theorem. Once a {\it catalogue} of properties and forces that define an object or physical system is established, the theorem allows one to select a complete set of Dimensionless quantities  or {\it Numbers} on which structure must depend. The  internal structure takes the form of a functionally defined manifold in the space of these Numbers. Simple and familiar examples are discussed by way of introduction. Correlations in  properties of astronomical objects can be sought  either through the constancy of these Numbers or between pairs of the Numbers. In either case, within errors, the functional dependences take on an absolute numerical character. As our principal application, we study a well defined sample of galaxies in order to reveal the implied Tully Fisher and Baryonic Tully Fisher relations. We find that $L\,\propto\,v_{rot}^4$ for the former and $M_b\,\propto\,v_{rot}^3$ for the latter, suggesting that these relations may have different  causal origins.

\end{abstract}
\newpage

\section{Introduction}

There has been an enormous effort to understand the message provided by correlations between properties of astronomical objects, known as scaling relations. The well known Tully-Fisher and Faber-Jackson (or fundamental plane) relations are classic examples. Some relatively current discussions concerning spiral galaxies are found in  \cite{Cour2007}, \cite{LMS2016}, and \cite{Pon2017}, but there are many others. The discussion centres on determining which pair of  quantities show the smallest scatter about the mean relation. Subsequently one seeks a way of minimizing the scatter by classifying galaxies according to a suitably chosen third (or more) variable. The importance of knowing hidden correlations is shown in the various derivations of the traditional Tully-Fisher relation based on the virial theorem and certain other assumed relations. These can be made to yield a slope in log (luminosity)-log (rotational velocity) space of  either $3$ or $4$, depending on which hidden variable is ignored \citep[e.g.][appendix B]{Cour2007}. 

A breakthrough was made in \cite{MSBD2000}, when the baryonic mass was shown to correlate better with rotational velocity (slope $4$) than the original Tully-Fisher correlation between luminosity and rotational velocity. This has been generally accepted \citep[e.g.][]{Pon2017}, but the exact slope remains dependent on various filters  used on the data \citep{Cour2007,Pon2017, Sch2017}.

The objective of such studies was initially to find a distance measure, for which the correlation of the luminosity with a directly measurable quantity is required. The Faber-Jackson relation for elliptical galaxies provides a similar luminosity measure in terms of the stellar velocity dispersion. Subsequent study revealed this relation to be a projection of a higher dimensional surface in a certain galactic property space, termed the `fundamental plane'. However the emergence of  such strong correlations between the macroscopic properties of galaxies has  also challenged simulations of galactic formation and evolution. 

 This paper does not pretend to answer such questions. However there is a starting point for the discovery of relations between galactic properties based on the Buckingham theorem from Dimensional analysis. This theorem does not seem to have been employed previously in an astronomical context, although it is generally used in physics \citep[e.g.][]{Hen2015}. It requires identifying Dimensionless {\it Numbers} (i.e. Dimensionless ratios), constructed by combining separate physical {\it measures} that are {\it intrinsic} to the object. It is  always possible to consider Dimensionless measures by simply scaling them by a fiducial measure (e.g solar metallicity, solar mass, solar luminosity or a reference speed), but these are just a form of {\it Units} and do not actually remove the Dimensionality \citep{Hen2015}. A price in kilo dollars is still money. Moreover they are not intrinsic to the system and do not appear in the {\it catalogue}.

 {\it Our approach initially discounts  arguments for correlations between physical properties, in favour of a macroscopic description of a galaxy.} This description requires agreeing on a satisfactory catalogue of galactic properties together with the forces that act.  Such a catalogue implicitly contains a mixture of physical and measurable choices. There is  the {\it ideal catalogue} which contains every quantity or force that can conceivably play a r\^ole in the nature of the object, many of which may be unmeasured or unmeasurable. There is more usefully for astronomy, the catalogue of measures or observations by which the object is defined. We call this the {\it measured catalogue}. In sum it really {\it is} the astronomical object. This catalogue will differ from observer to observer in general depending on the assembled observations. It will also evolve as more data become available.   More practically, one can construct the {\it hybrid catalogue} which will contain the best possible combination of measured and ideal parameters, as can reasonably be ascertained from current knowledge.  Examples are given in the next section.
 
 Once a catalogue is adopted either ideal or known, the Buckingham theorem \citep[e.g.][]{Hen2015}  allows a {\it rigorous and complete}  calculation of the Numbers on which the  structure of the object must depend. If the catalogue contains $n$ objects, which have among them $m$ independent Dimensions, then a set of $n-m$ Dimensionless Numbers made from the catalogue objects is a complete set. Normally, astronomical objects will involve mass, length and time so that $m=3$, implying $n-3$ Numbers in the complete set. The special case whenever $n-m=1$ is important because one can conclude that the single Number is a constant. This gives a strict relation between the Dimensional quantities that comprise it. 
 
  In general these  Numbers form a multi-dimensional space and the relation  between them must be projected onto a sub-space for practical use. This is comparable to the fundamental plane for elliptical galaxies, but in this case the axes are pure Numbers. The  bivariate projections normally require neglecting the remaining Numbers for various reasons, which reasons reintroduce physical argument of a general kind. Observably constant Numbers also  establish  correlations between their constituent quantities. Before beginning the astronomical study, we proceed in the next section to illustrate the argument by simple examples.

\section{Physics by the Numbers: Good and Bad}

We consider some  illustrative examples from engineering and experimental physics, followed by particularly simple, if extreme, astronomical phenomena.

\subsection{Pipe Flow}

Let us suppose that we have a view of  horizontal pipe flow that corresponds somewhat to the observational reality of an astronomical object. We can only detect the emission of fluid from the delivery end of the pipe and it appears to be in a steady state. The observations include collecting some of the substance and measuring its viscosity $\eta$ and density $\rho$, which we find to be incompressible. We also measure the `luminosity' of the pipe as the emitted mass per unit time, $L$, and the radius of the aperture, $a$. This allows us to infer the mean efflux velocity as $\bar v\equiv L/(\rho a^2)$. 

 Let us suppose that unlike the  pipeline engineers, we are not instrumented to detect the velocity or pressure profiles in the cross section and, because the pipe  is horizontal on the Earth, we do not regard gravity as a driving force\footnote{The line must be short enough to avoid the curvature of the Earth}. Dependence of the flow on radius in the pipe, the gravitational effect on the pressure profile, and the temperature of the fluid would all be part of the {\it ideal catalogue}, which depends in part  on unobserved physical principles. We can not vary the radius and transverse gravitational field,  but we expect their importance to vanish in the average behaviour over the cross section. We {\it can} measure the temperature $T$ in the collected material.  It should be noted that, in an {\it ideal catalogue}, the constants that characterize
 {\it how} density and viscosity depend on temperature would replace the constant values of density and viscosity.
This is  beyond our direct measurement capabilities if we observe at a given operating  temperature.  Otherwise we might rely on `meta measurement' that is on theoretical principles, but  here we assume that our measurements are all at the same temperature. 

Because of the measured mass flux, viscosity, density, and radius we can infer by physical principles that an incompressible fluid in a steady flow must experience a constant  mean drag force per unit area of order $\eta L/(\rho a^3)\equiv \eta \bar v/a$ by traditional Dimensional analysis. There must consequently, again by Dimensional analysis, be a constant driving force per unit area at each cross section of order $\Delta p\equiv a|dp/dz|$. Here $z$ is the coordinate along the pipe, but the delivered behaviour does not depend on its value (unlike the starting behaviour) and so $a$ is the only relevant scale.  The pressure gradient, $|dp/dz|$, is an element of `meta measurement' that strictly is not directly accessible to us. However we choose to include it because the relevant physical principle is clear and relatively unique. 

We can now assemble a {\it hybrid catalogue}, $HC$, representing our end of the pipe flow as a mixture of {\it measured} and {\it ideal} catalogues combined with intuitive physical principles. This  catalogue represents a kind of positivist view of the pipe function  and becomes 
\be
HC\equiv \{\eta,\rho,a,L,\Delta p,kT\},\label{eq:Hybrid1}
\ee
where we have introduced Boltzmann's constant $k$ so that each member of the catalogue can be expressed in terms of one of the three independent mechanical Dimensions, namely length, mass and time.
There are six quantities in our catalogue and three independent dimensions. Hence by the Buckingham theorem  (sometimes called the `Pi' theorem \citep[e.g.][]{Hen2015} there are three independent Numbers that must be connected functionally in any statement about the system behaviour. These Numbers are not {\it unique}, but they are {\it complete}, so that any other choice reduces to any particular complete choice. {\it One can also transform a Number by any useful function without affecting the completeness of the set. For example reciprocals, roots, logs et cetera, do not change the essence of a Number.} 

In the present example we can choose a particular complete choice  of Dimensionless ratios or Numbers as 
\be
Re\equiv \frac{L}{\eta a},~~~~~RF\equiv \frac{\rho\Delta p a^2}{\eta^2},~~~~~U\equiv \frac{ kT/a^3 }{\rho \bar v^2}\equiv \frac{\rho kT a}{L^2}.\label{eq:Numbers1}
\ee

The first Number $Re$  is the Reynolds number of the flow, the second number $RF$ is an estimate of  the  squared ratio of the driving force per unit area to the drag force per unit area (i.e multiply top and bottom by $\bar v^2/a^2$ to get $\rho(\bar v)^2\Delta p$ and $(\eta\bar v/a)^2$).  Finally $U$ is the ratio of the thermal energy per unit volume to the kinetic energy per unit volume in a section of the pipe.  We therefore have that, for the end point observer, any statement regarding properties of the pipeline delivery system is contained in the functional relation  
\be
f(Re, RF,U)=0 , \label{eq:Pipe1}
\ee
  
Let us first assume that we are strict positivists and so we exclude the meta measurement $\Delta p$. Then only two Numbers remain as arguments in equation (\ref{eq:Pipe1}). In fact we expect to be able to invert this equation \citep[e.g.][]{Hen2015}  to obtain the relation $U=F(Re)$ or explicitly 
\be
 \frac{\rho kT a}{L^2}=F(Re).\label{eq:Pipe2}
 \ee
 {\it This is the limit of the general argument in this case}. We conclude only that if we observed with the same philosophy many different end pipes, then the functional form could be found `observationally'. This would then relate {\it in a Dimensionless equation} the individual physical  quantities occurring in the positivist description of pipeline delivery. This is necessarily a causal relationship (within errors) rather than just a correlation, whenever there are only two Numbers.  If there were only a single Number, then it should be a constant.
  
 One can proceed approximately with the full relation (equation \ref{eq:Pipe1}) in various ways. By looking at an ensemble of pipe flows, one of the Numbers may be found to be a function of the other two (always within measurement errors). This defines a surface in the space of the three Numbers, which constitutes  a `fundamental' causal constraint on the physical quantities comprising the Numbers. 
 
 It may be that one or more of the Numbers does not vary widely over the pipeline universe, that is each Number is nearly constant over the ensemble. In that case  each Number is unlikely to be an essential distinguishing characteristic over the ensemble and one is justified in ignoring this Number in lowest order. The actual small variation in any Number ignored, will be a source of scatter in the remaining relations (in addition to measurement error). The fact that any Number is approximately constant is already a simple causal constraint on the physical quantities that comprise it. When only one independent Number exists then, as mentioned in the introduction, it must be constant \citep[e.g.][]{Hen2015} and so supplies a necessary constraint.
 
 Finally if one Number is enormously small (or if large then the reciprocal) compared to all others, then a simple MacLaurin expansion in this Number may clarify things. If for example $U$ is much smaller than $Re$ or $RF$ in equation (\ref{eq:Pipe1}), then by a MacLaurin expansion the surface reduces to the bivariate relation of equation $f_0(Re,RF)=0$ {\it in lowest order}.

 
  
 We return now to the hybrid catalogue and its consequence summarized in equation (\ref{eq:Pipe1}). Again assume either that for sufficiently cold pipe flows $U\ll 1$, or that $U$ is approximately constant over the pipeline ensemble. In either case  we infer that (our functions are assumed to be invertible)
 $RF=f(Re)$, which is explicitly 
 \be
 \frac{\rho\Delta p a^2}{\eta^2}=f(Re).\label{eq:Pipe3}
 \ee
 
 Once again one has to observe many pipe line operations to measure the function of $Re$. This has been found to be $C \times Re$ where $C$ is a universal constant. One might guess at this by realizing that equation (\ref{eq:Pipe3}) applies to all pipe lines in steady state operation. Because there is no acceleration, one might expect the dependence on inertia ($\rho$) to cancel out in any structure relation. This occurs with a linear function of $Re$ on the right, on recalling the definition of $L=\rho\bar v a^2$. In any case adopting this result for $f(Re)$, we obtain explicitly from the definitions 
 \be
    L=\frac{1}{C}\frac{\rho\Delta p a^3}{\eta}\equiv \frac{1}{C}\frac{\rho |dp/dz| a^4}{\eta} \label{eq:Pipe4}
 \ee
This last result is the Hagen-Poiseuille `law of the pipe', if $C=16/2\pi$ \citep[e.g.][]{Hen2015}.  That the unknown numerical constant should be of order unity is typical of these arguments. Equation (\ref{eq:Pipe4}) can also be read as yielding the pressure gradient necessary to produce a given mass flux  in a pipe of radius $a$ with a given density and viscosity. 

There is one other way in which $U$ may not appear, which is of particular relevance to astronomy. It could be that even for the end point of the pipe line 
we are not able to measure all relevant quantities {(say, either star formation rate or stellar/gas velocity dispersion in astronomy)}. In that case our data set only provides an {\it Effective } catalogue, which will in general also be hybrid. For example if the temperature of the fluid is unavailable to us for any reason, then the {\it Effective} catalogue for the pipe line ensemble is the catalogue (equation~\ref{eq:Hybrid1}) minus $kT$. Then we infer based on this effective catalogue the result (equation~\ref{eq:Pipe3}). However now we may be unaware of the scatter in relation  (\ref{eq:Pipe3}) as a result of ignoring $U$. If the relation is nevertheless well defined, we would infer the unimportance (i.e. constancy, or  very small  values) of the missing Number(s). If indeed $U$ is constant, then we have the unusual relation for an ensemble of varying temperature (although we ignore density and viscosity variations that imply unrecorded Numbers so our catalogue is indeed Effective) 
\be
kT=U\times \rho a^3 \bar v^2.
\ee

 Note that the above results could not have been found from dimensional analysis alone and are unlikely to have been discovered via trial-and-error correlations between properties.
We have dwelt in detail on this well-known example because it sets the pattern for  subsequent examples in this paper.  It is a theory section in disguise.   As suggested above, even if the catalogue is {\it Effective}, or {\it Hybrid} or incomplete, this approach provides a {\it systematic} method of achieving a physical understanding of the relations between the members of the catalogue.  We have emphasized the decisions to be made and the absolute nature of relations between Dimensionless Numbers (e.g. equation~\ref{eq:Pipe4} with $C$ universal). 

We have not so far discussed the disadvantages that arise in experimental practice. These are due to the inevitable errors associated with individual  measurement. By calculating the Dimensionless Numbers these errors are combined to become larger than the error of each quantity. This can be mitigated to a certain extent by a careful choice of the Numbers, but the errors can still lead to substantial scatter and must be propagated carefully.   On the other hand, a physical correlation found through trial and error may show a scatter outside of the error bars of the (two) properties being correlated due to a missing `3rd parameter'.  While including the 3rd parameter may reduce the scatter, the error bars must necessarily be larger in any case because of the now-included error in that 3rd parameter. An astronomical example is the Period-Luminosity Relation for Cepheids, the 3rd parameter being colour.  Thus, the error bars in our method are only larger in so far as we initially retain all parameters that are actually required to describe the phenomenon, together with their error bars. 

When the final statement from the Buckingham theorem involves more than two  important Numbers (e.g. three as in equation~\ref{eq:Pipe1} or more), then additional scatter when plotting the mutual dependence of any two Numbers will be due to the influence of the others.  There is  art plus trial and error involved in choosing which Numbers are important and which are of lesser importance. The final conclusion may or may not differ substantially from traditional techniques. However, the process ensures that no important parameters from the catalogue or their relations are missed. Paradoxically, both those  Numbers that vary widely and those  that are nearly constant are of interest. It all begins with the choice of catalogue, ideal, hybrid and usually effective that requires the application of  physical insight and the recognition of incomplete data. 

In the next sections we indulge ourselves a little by looking at extreme astronomical objects with the same methods and philosophy.  Finally we arrive at a more substantial example for an ensemble of spiral galaxies.

\subsection{Black Holes}

Black holes are perhaps the best objects for which a positivist approach applies. Unlike stars, the interior and surface  physics are not essential to their external nature. 

We detect the existence of black holes simultaneously with their mass $M$, by detecting their dynamical influence on surrounding stars. We suspect gravity, in a non-stellar, singular limit, is their cause. Hence we must include $G$ in their catalogue as the force coupling constant. This represents the inclusion of a physical principle so that the final black hole catalogue will be hybrid. The dynamical influence of the black hole depends on the distance from it, so we  add a spherical radius $r$ to the catalogue that may be observable. The observational phenomenon of lensing and micro-lensing tells us that the black hole interacts with light. Consequently the vacuum electromagnetic wave speed  $c$ should also be added. From the effects on gas being accreted we can expect to detect the rotation of the black hole. In fact we would expect its angular momentum to be the real factor, and so we add  to the catalogue a specific angular momentum (we already have the mass) $a$ ($a$ in this example is {\it not} a linear scale as it was in the introductory example). We are also assuming another physical principle, namely that the black hole is axially symmetric. In spherical coordinates based on the rotation axis, this will introduce the polar angle $\theta$ relative to the axis of rotation into the catalogue. We have not yet measured quantum effects so we omit Planck's constant $h$, but this omission probably creates an {\it effective} black hole catalogue.

According to the arguments of the  preceding paragraph, we should construct the black hole catalogue (effective hybrid) as 
\be
BHC\equiv\{M,a,r,c,G,\theta\}.\label{eq:BH1}
\ee
There are {\it six} defining  quantities and the Buckingham theorem requires that statements about Black Holes should take a functional form in terms of {\it three}  Numbers because there are only three independent Dimensions (six quantities minus three independent Dimensions of Length, Mass, and Time).  These may be taken to be $\{E,L,\theta\}$ where 
\be
E\equiv\frac{rc^2}{2GM}\equiv \frac{r}{r_g},~~~~~L\equiv\frac{a}{cr}~~~~ and ~~~\theta,\label{eq:BH2}
\ee
 or any combination of these three. Angle is already a Dimensionless Number, and we have cheated slightly in the interests of familiarity by introducing the factor $2$ to define $r_g=2GM/c^2$. This is permitted by the theorem but it can not be foreseen without prior knowledge. Fortunately we will see that this does not really change the argument in any substantial way. Any statement about the observed structure of the black hole now takes the form
 \be
 f(E,L,\theta)=0,\label{eq:BH3}
 \ee
 and without more information this manifold in three Number space is the rigorous result of the argument.
 
  Assuming the extra principle of spherical symmetry (which requires zero rotation) there is only one number $E$ and this is required to be constant by the Buckingham theorem. Hence in that limit we infer that $r=C\times 2GM/c^2$, with $C$ a universal constant for all static, spherical black holes. We interpret this value for a given object as the radius of the black hole, which is thus determined by our arguments to within a numerical factor $C$. On general grounds we can expect $C=O(1)$, but in fact we know that it is exactly one, but that result is beyond the positivist approach. Of course when only one Number is available the argument reduces to standard Dimensional analysis, but this systematic approach reveals the limitations. 
 
  To go beyond equation  (\ref{eq:BH3}), more physical principles must be introduced, which introduce a kind of meta measurement. We associate gravity with the local metric and differential geometry of space-time. More particularly we associate it with a Riemannian measure of space-time $ds$. This adds to our effective hybrid catalogue $ds$ and $cdt$ and hence two new  Numbers $ds/r_g$ and $cdt/r_g$ or any functions of these.  The Buckingham theorem, generalized from equation(\ref{eq:BH3}) and  the implicit function solved for $ds^2/r_g^2$ can be written as 
  \be
  \frac{ds^2}{r_g^2}=f(E,L,\theta,\frac{cdt}{r_g}).\label{eq:BH4}
  \ee
  However we know the general form of an axially symmetric Riemannian metric. Together equation (\ref{eq:BH4}) and the assumed differential geometry implies that the functional  equation (\ref{eq:BH4}) takes the form
  \bea
  \frac{ds^2}{r_g^2}&=&g_{00}(E,L,\theta)\frac{c^2dt^2}{r_g^2}-g_{11}(E,L,\theta)\frac{dr^2}{r_g^2}-g_{22}(E,L,\theta)\frac{r^2d\theta^2}{r_g^2}\nonumber\\
  &-&g_{33}(E,L,\theta)\frac{r^2\sin^2{(\theta)}d\phi^2}{r_g^2}+L~g_{03}(E,L,\theta)\frac{cdt~d\phi}{r_g}.\label{eq:BH5}
  \eea

In equation (\ref{eq:BH5}) we have adopted the sign convention $-2$ and the $g$ functions are both positive and Dimensionless. The factor $L$ in the final term is simply to allow the limit $L\rightarrow 0$ to be taken easily. The Numbers $E$ and $dE$ could replace $r/r_g$ and $dr/r_g$ respectively, which is more faithful to our argument but less familiar.

In the spherically symmetric limit $L\rightarrow 0$ and $\theta$ does not appear in the Dimensionless functions, so that  equation (\ref{eq:BH5}) becomes 
\be
ds^2=g_{00}(E)c^2dt^2-g_{11}(E)dr^2-r^2(d\theta^2+\sin^2{(\theta)}d\phi^2).\label{eq:BH6}
\ee
This Schwarzschild-like space-time can be made more explicit by adding the principle that we should have Minkowskii space (locally in a flat Universe) as $r\rightarrow \infty$. We expand about $1/E\equiv r_g/r=0$ to find $g_{00}\rightarrow 1+(g_{00})_0\frac{r_g}{r}$ and $g_{11}\rightarrow 1+(g_{11})_0\frac{r_g}{r}$. Choosing the pure numbers $(g_{00})_0=-1$ and $(g_{11})_0=+1$ gives a reasonable approximation to the Schwarzschild metric. Had we not adopted the factor $2$ in the definition of $r_g$ then these values would be simply $\mp 2$.  Both undetermined constants are $O(1)$. The result is exact for $g_{00}$ because the spherical symmetry removes any tidal effect, but it is only a crude first order expansion for $g_{11}$ that can not account for the eventual coordinate singularity. 

The general result (equation \ref{eq:BH5}) does not lend itself to explicit results except for expansion in small $L$. Nevertheless the form agrees with that of the Kerr metric in Boyer-Lindquist coordinates and would be of some help in solving the Einstein equations.

\subsection{Cosmology}

The Universe is an excellent example of an object that is known by the `Numbers', although it is interpreted according to the Friedmann-Lema\^itre  (Robertson-Walker) models of the differential geometry. There is only one currently  observed Universe (by definition) so that we can not test our relations over an ensemble. We might do this by looking at the set comprising the Universe at different times (red shifts $z$), but {here} we restrict ourselves to the observed structure at the present epoch (small  $z$). 
A typical catalogue  for the low red shift Universe is  
\be
CC\equiv\{ G,c,H_o,\Lambda_o, \bar\rho,kT_b,\frac{\delta T_b}{T_b},a_o,t_o\}.\label{eq:Cos1}
\ee
This is a hybrid catalogue since we use physical principle to include $G$ and $c$; while the Hubble constant $H_o$, cosmological constant $\Lambda_o$, background temperature $T_b$, mean density $\bar\rho$, and fluctuations in the background temperature  $\delta T_b/T_b$ are observed.  The subscript $o$ indicates current epoch but the analysis applies at any epoch. We have also included the Mond  \citep{Mil1983} acceleration $a_o$ for illustration.  The age of the Universe  is $t_o$. The fluctuation spectrum and structure correlations are ignored as reflecting  internal details rather than gross behaviour. Planck's constant is omitted for similar reasons.

We expect to construct six Numbers  ($9 -3$) from this catalogue and we take them to be
\be
K\equiv \frac{G\bar\rho}{H_o^2},~~~V\equiv \frac{\Lambda_o c^2}{H_o^2},~~~F\equiv \frac{\delta T_b}{T_b},~~~A\equiv \frac{a_o}{c^2\sqrt{\Lambda_o}},~~~\tau\equiv \frac{kT_b}{\bar\rho c^2}\Lambda_o^{3/2}~~~~T\equiv H_o t_o.\label{eq:Cos2}
\ee

For a single object, the one Universe at the current epoch, {\it all} of these Numbers are constant. There is a statement about the structure of the Universe that must, according to the theorem, take the rigorous form
\be
f(K, V,A,T,F,\tau)=0,\label{eq:Cos3}
\ee
which implies some relation between the constant arguments.  
However any reasonable estimate makes $\tau$ singularly small and even $\delta T_b/T_b=O(10^{-5})$ while other quantities are $O(1)$ according to current estimates.  We assume  a MacLaurin expansion of equation (\ref{eq:Cos3}) to remove these Numbers in zeroth order and consequently any  physical relation between the constants should take the form 
\be
f(K,V,A,T)=0.\label{eq:Cos4}
\ee
The same relation should hold at any epoch  until $H_o$ and $\Lambda_o$ became very large. Measuring these Numbers at very different epochs might reveal the relation or projections of it.

 At any epoch equation (\ref{eq:Cos4}) now predicts a relation between the current numbers $K,V,A ,T$. At the current epoch however, there is only one observed Universe in which $K$, $V$, $A$  and $T$ have constant values. The Number $T=constant$ yields the expected 
expression for the age of the Universe $t_o=T/H_o$, where $T=O(1)$ if the Planck satellite values for $t_o$ and $H_o$ are used (\cite{Planck2018}). The other constant Numbers are also interesting. 

From $K=constant$ we obtain $\bar\rho=(H_o^2/G)K$, which is $0.7K\times 10^{-28}~g~cm^{-3}$ if $H_o=67 ~km ~s^{-1}~Mpc^{-1}$ as per Planck. It becomes $0.83 K\times 10^{-28}~g ~cm^{-3}$ if $H_o=73$, in the same Units. We might take $K=3/(8\pi)$ as per Friedmann-Lema\^itre theory, but `a priori' we expect it to be closer to one (recall $C/2\pi$ in Hagen-Poiseuille flow).

From $V= constant$ we obtain $\Lambda_o=(H_o^2/c^2)V$, which is either $0.53V\times 10^{-56} cm^{-2}$ or $0.62V\times 10^{-56}~ cm^{-2}$ depending on whether the small or large value of $H_o$ is chosen. With $V=O(1)$, each of these values is within a factor $2$ of the Planck value of $1.1\times 10^{-56}~ cm^{-2}$. Either one of the constants $K$ and $V$ could be considered as giving a value for $H_o$, if either $\bar\rho$ or $\Lambda_o$ were better determined than $H_o$.

Combining $K$ and $V$ to eliminate $H_o$ gives the `dark energy' density compared to the total matter density as 
\be
\frac{\Lambda_o c^2}{G\bar\rho}=\frac{V}{K},\label{eq:Cos5}
\ee
which gives $V/K=7.9$ using the Planck value for $\Lambda_o$ and the critical matter density. If the value $\Lambda_o/(8\pi)$ is used to calculate the vacuum energy density, then $V/K\approx 2.2/\pi$ (i.e. of $O(1)$) in order to fit the Planck result for $\Omega_\lambda$.

The Number $A$ only exists if one believes in a fundamental acceleration constant $a_o$ at the present epoch, according to the `Mond'  \citep{Mil1983} hypothesis. With $A$ non zero but constant we evaluate this quantity as $a_o=A\times c^2\sqrt{\Lambda_o}$ which is $\approx 0.9A\times 10^{-7}~ cm ~s^{-2}$ using the Planck value for $\Lambda_o$. 
This is nearly a factor of ten larger than that deemed optimal \citep[$ 1.2\times 10^{-8}~cm~s^{-2}$,][]{Mil1983} by fits to galaxy rotation curves, if $A=O(1)$.  One can bypass $\Lambda_o$ by combining $V$ and $A$ to write $a_o=A\times\sqrt{V} H_o c$, for which the smallest value is $\approx 0.65~A\sqrt{V}\times 10^{-7}~cm~s^{-2}$, about five times too large. 

This example emphasizes that these  constant Number relations do not indicate causality. unless they are really of $O(1)$.  Otherwise they are purely a result of the free choice of Units together with Dimensional analysis \citep{Hen2015}. The notion of causality only arises when a real functional form can be found in the implicit relations  of the type (equation~\ref{eq:Pipe1}), (equation~\ref{eq:BH3}) or (equation~\ref{eq:Cos4}) or the equivalent.  Such relations are found by examining an ensemble of similar systems {\it or, in the case of an evolving single system, at different times}.   Our hybrid positivist procedure is independent of the FL-RW evolution theory, although just as for black holes the two could be used together to create the necessary form of the differential geometry. The Numbers in our catalogue of the evolving Universe would simply drop the subscript $o$.  Our one causal result  (e.g. equation~\ref{eq:Cos4}) suggests that it is not surprising that  $a_o$ should be related to $\Lambda_o$ or $H_o$ \citep[cf. ][for another view]{AS2018}.

\section{Spiral galaxies and the Tully-Fisher relation}
\label{sec:spiral}

In this section we apply our arguments to a highly selected sample of galaxies \citep{Pon2017,Pon2018}
which was studied to investigate the Baryonic Tully-Fisher relation (BTFr). This data set was adopted because it represents comprehensive measurements of a wide range of parameters in a consistent manner.

The Dimensional analysis view of a galaxy is necessarily coarse.  It reduces to a catalogue of  forces and properties, the latter being a set of observables. We have one great advantage concerning the forces acting on galaxies in that we expect gravity to be  dominant.  Newton's constant $G$ must  therefore appear in the catalogue. The  electrically conductive property of galaxies due to cosmic rays allows us to ignore electrostatic fields, but the magnetic field may play a r\^ole. Taking the conductivity to be infinite allows us to discuss only the magnetic field $B$ and possibly the induced electric field $-{\bf v}/c\wedge {\bf B}$.  However the latter field is small if the velocity is small relative to the speed of light $c$. 

As for the observable properties, we choose first the total  dynamical mass $M$, the baryon mass $M_b$ and the stellar mass $M_*$.  The  central black hole mass will be ignored as being small compared to the total mass. This would not be true if we were considering  primarily the spiral bulge. The luminosity in some band of choice will be taken as $L_\nu$ and a characteristic scale will be labelled by $R$. It is relatively unimportant from the Dimensional analysis point of view as to which band and scale  are actually chosen, so long as the choice is done in the same way for all galaxies. 

The velocity associated with the spiral galaxy will be divided into two parts. There is the rotation velocity $v_{rot}$, which again need only be measured in the same way for all galaxies, but which we normally associate with the flat part of the rotation curve in HI, $v_{flat}$. In addition there is a turbulent component reflecting velocity dispersion on the stellar scale \citep[e.g. ][]{Tam2009} which I will designate $\sigma$. This is measurable in galaxies  for example, by the line width in optically thin H alpha.  When the turbulent component is not observed for each galaxy, as is the case for the sample of this section, we introduce the star formation rate $S_*$. Because we do not have both, our catalogue will be `effective'. The actual relation between star formation and velocity dispersion has been the subject of much debate \citep[see e.g.][and references therein]{MTK2015}, but some relation seems to exist. We would need to add the Toomre Number $Q$ (see below) to our sample in order to study this relation. 

There is also evolution of the galaxy  as summarized in the redshift  distribution $\{z\}$. We suppress $z$ here from the outset because we observe nearby  evolved galaxies for which $z$ does not vary widely over the sample. Introducing gravity and the magnetic field requires the application of physical principle (meta measurement) as indeed would be also the inclusion of $z$. Our catalogue will then be an `effective hybrid' mix of observations and physical principles.

This survey of properties gives our coarse description of a spiral galaxy in the sample of this section, as the catalogue 
\be
SC1\equiv\{G,B,M,M_b,M_*,L_\nu,v_{rot},S_*,R\},\label{eq:spiralgalaxy}
\ee
from which we may infer statements about galactic structure. As before, Dimensionless structure is restricted by the Buckingham theorem  \citep[e.g.][]{Hen2015} which tells us that from the catalogued `nine' we can form only six independent, {\it Dimensionless} Numbers \footnote{{As before,} the three independent Dimensions with arbitrary Units allow this reduction.}. These six Numbers are not unique in form as we have seen, but any independent six are complete. That is, {\it there are no other Numbers independent of these six.}  For the purposes of this analysis we choose the Numbers to be
\bea
\mu_*&\equiv& \frac{M_*}{M},~~~~~\mu_b\equiv\frac{M_b}{M},~~~~~{\cal L}\equiv \frac{L_\nu}{S_*v_{rot}^2},~~~~~{\cal S}\equiv \frac{GS_*}{v_{rot}^3},\nonumber\\
{\cal V}&\equiv&\frac{GM}{Rv_{rot}^2},~~~~~{\cal M}\equiv\frac{R^3B^2}{Mv_{rot}^2}\label{eq:spiralnumbers1}.
\eea

The expected functional statement of structure takes the form 
\be
F(\mu_*,\mu_b,{\cal L},{\cal S},{\cal V},{\cal M})=0.\label{eq:basicR}
\ee
which should include any Dimensionless statement concerning  galactic properties, given our catalogue. It is a five dimensional relation in a six dimensional space of Numbers, without further reduction. We emphasize that this simple step contains {\it all} of our {\it  rigorous} analysis. 

Should the velocity dispersion $\sigma$ be part of the galaxy description in our sample, then the Number $Q\equiv \sigma/v_{rot}$ would appear. However in this sample $\sigma$ is taken constant for all galaxies in the sample \citep{Pon2016} so that it does not contribute to the variability in the ensemble, although the Number $Q$ does. It should be noted that at galactic radii determined by the HI distribution, the squared epicyclic frequency is $GM/R^3 $ in terms of the dynamic mass  $M$ and the HI radius $R$  \citep[e.g.][]{BT2008}.  The  local Toomre condition for stability of the stellar disc, becomes essentially  equal to $Q$ according to 
\be
\frac{Q}{3.36\sqrt{\cal V}}>1,\label{eq:Toomre}
\ee
where we have set the surface density equal to $M/R^2$.  Although $Q$ is clearly an important `Number' for a sample including dispersion, we suppress it  here nevertheless  as not varying sufficiently widely over the sample. See also  the discussion of `filters' that follows. 

To make further progress we must have a means of determining which (if not all) subset of the Numbers is most significant.  One filter is to suppress a Number which is constant (within errors) over the ensemble. The virial Number ${\cal V}$ is usually held constant as a means of calculating the dynamical mass $M$ of the galaxy. If we take this literally, then we may disregard ${\cal V}$ as a significant variable over the galaxy ensemble. This says explicitly that  by definition $M={\cal V}~Rv_{rot}^2/G$. 

\begin{figure}
\begin{tabular}{cc} 
\rotatebox{0}{\scalebox{0.35} 
{\includegraphics{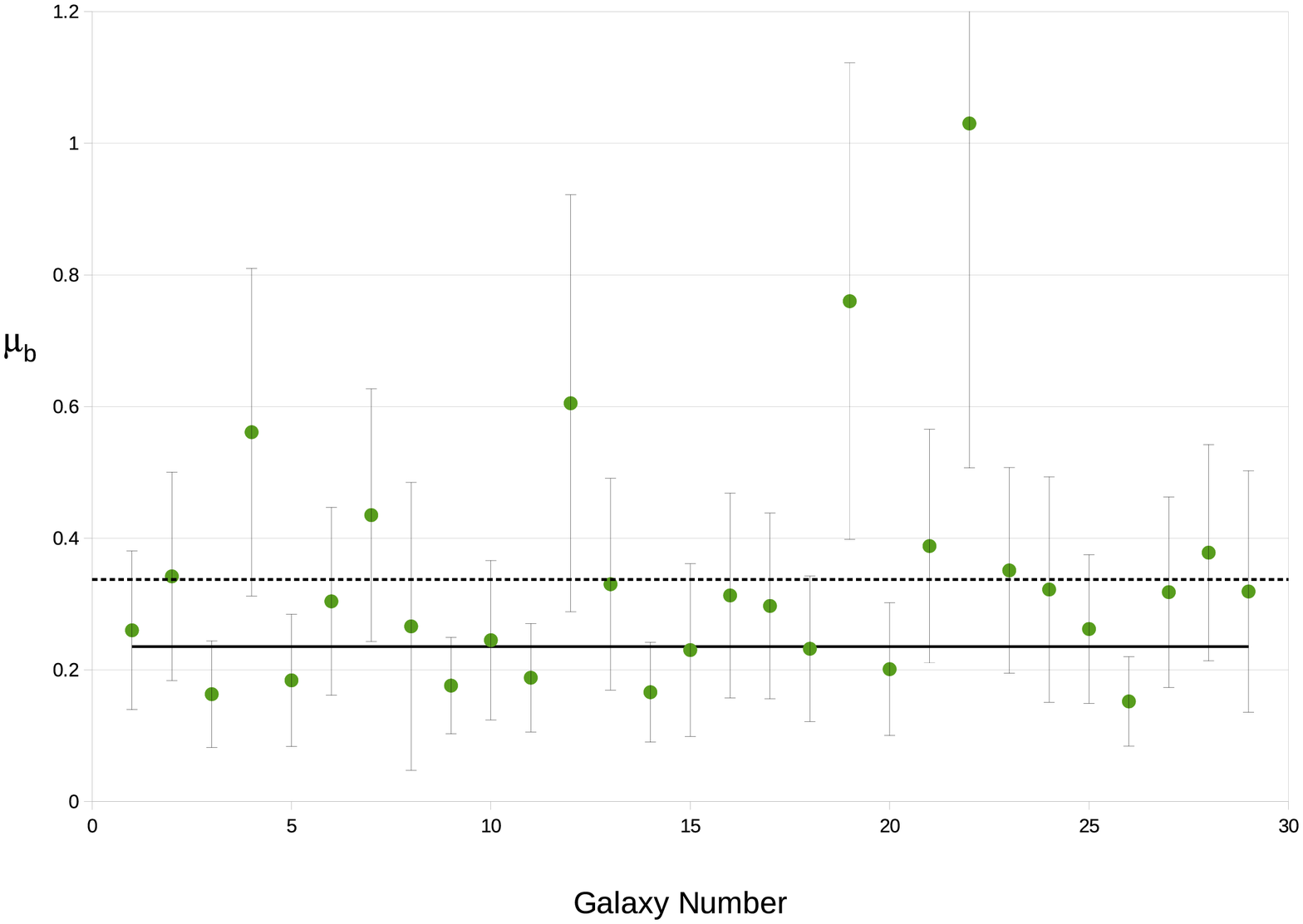}}}
\rotatebox{0}{\scalebox{0.35} 
{\includegraphics{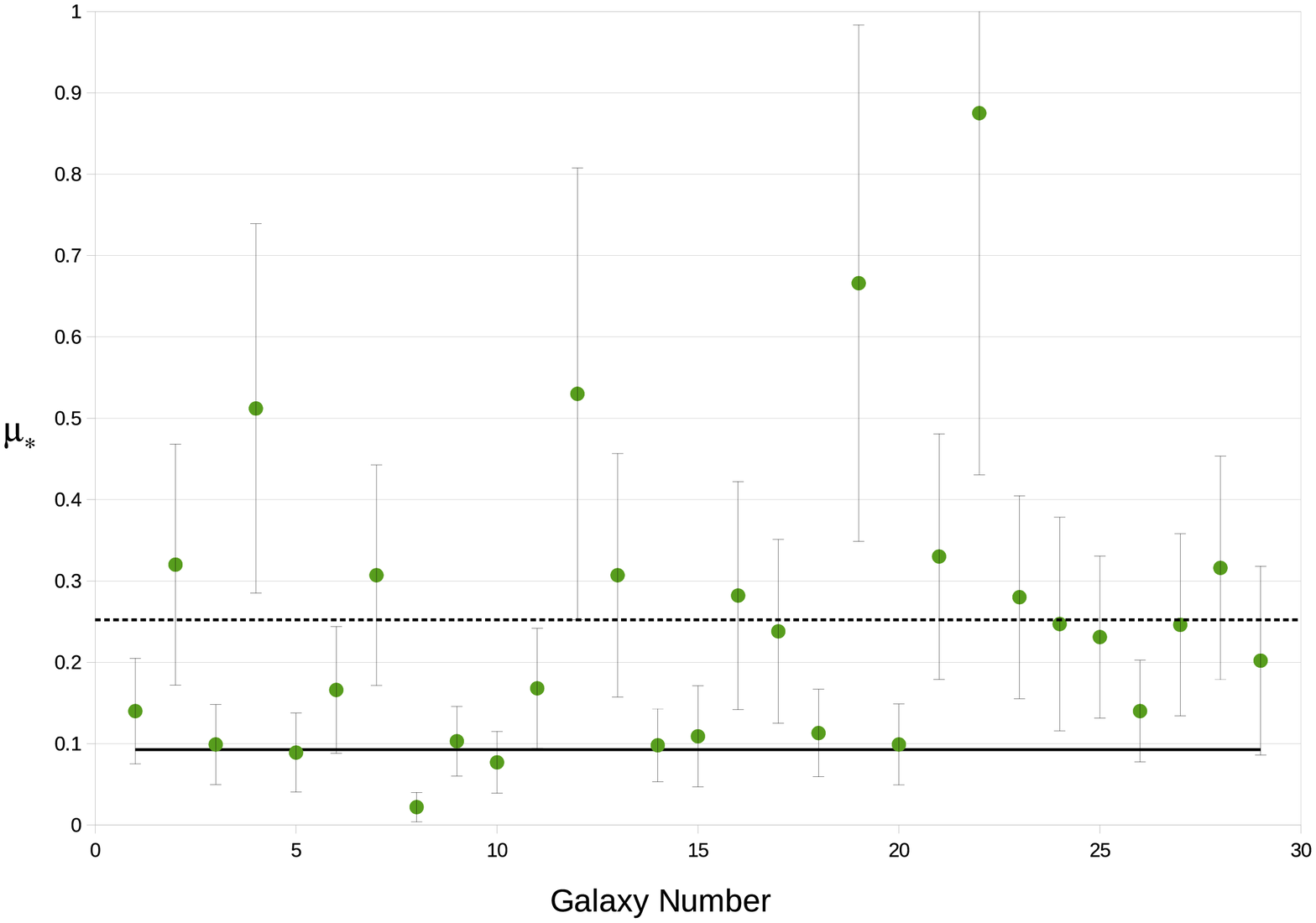}}}\\
\end{tabular}
\caption{On the left we show the values of the baryonic mass fraction of the dynamic mass for all of the galaxies in the sample. The solid line shows the weighted mean value of $0.235\pm 0.023$. The dotted line shows the unweighted mean with value $0.337$. Each point is treated equally in the unweighted case. On the right of the figure we show the same values for the stellar mass fraction of the dynamic mass (without gas mass). The weighted mean value is $0.093\pm 0.011$ and the unweighted mean value is $0.252$. } 
\label{fig:massfractions}
\end{figure}

It is interesting to remark  at this early stage that a recent paper \citep{Metal2018} has argued for a linear relation between the maximal HI radius and the corresponding rotational velocity ($v_{flat}$ for the current sample) over an ensemble of galaxies. They call this the `clock relation'. This relation is equivalent to a constant characteristic  global  density  over the ensemble, that is $M\propto R^3$.  This constant density suggests in turn that $ \mu_b=M_b/M\approx constant$, because  then the implied $M_b\propto v_{rot}^3\propto R^3$ becomes with the `clock relation' the Baryonic Tully-Fisher relation (BTFr) yielded by the present sample \citep{Pon2018}.  We verify constant $\mu_b$ over the sample in figure (\ref{fig:massfractions}).    

Another indication that a Number may be filtered out is if it has a value of $O(1)$, or if its variation is of $O(1)$, in the presence of very large or very small  (inverse being very large) alternate Numbers that are widely varying over the sample. It transpires that in the sample of this section $\mu_*$, $\mu_b$ , $Q$ and ${\cal M}$ are all of $O(1)$ with variations of the same order. We show  plots of $\mu_b$ and $\mu_*$  over the sample in figure (\ref{fig:massfractions}). The dynamical mass is calculated by taking ${\cal V}=1$ and the radius required is taken to be $D_{HI}/2$  as given for the same sample in \cite{Pon2016}.

We see from the figure that, but for several outliers with large errors, $\mu_b$ has a well defined weighted mean, and the weighted mean of $\mu_*$ is only slightly worse. In addition to confirming our reason for filtering out these ratios, the figure indicates that to good accuracy both $M_b$ and $M_*$ are proportional to the dynamic mass $M$, although the baryonic mass proportionality is better defined.
 
The magnetic Number ${\cal M}$ is calculated using $B=10^{-5}$ Gauss and is also $O(1)$ for our sample, and  the Number $Q$ only varies as $1/v_{rot}$.  Thus we also filter these Numbers out. This leaves only the Numbers ${\cal L}$ and ${\cal S}$ to use in the analysis of our sample as a first approach. We expect only a weak scatter over the sample due to the existence of the other Numbers. This is especially so because the Number ${\cal L}$ is characteristically $10^{6}$ but varies over three or four orders of magnitude.
The Number ${\cal S}$ is characteristically $10^{-5}$ but varies over a similarly wide range. We therefore conclude that we should find the relation (inverting $F({\cal L},{\cal S})=0$)
\be
{\cal L}=f({\cal S}),\label{eq:LS}
\ee
as the best way of distinguishing members of the ensemble physically. 

\begin{figure} 
\begin{tabular}{cc} 
\rotatebox{0}{\scalebox{0.50} 
{\includegraphics{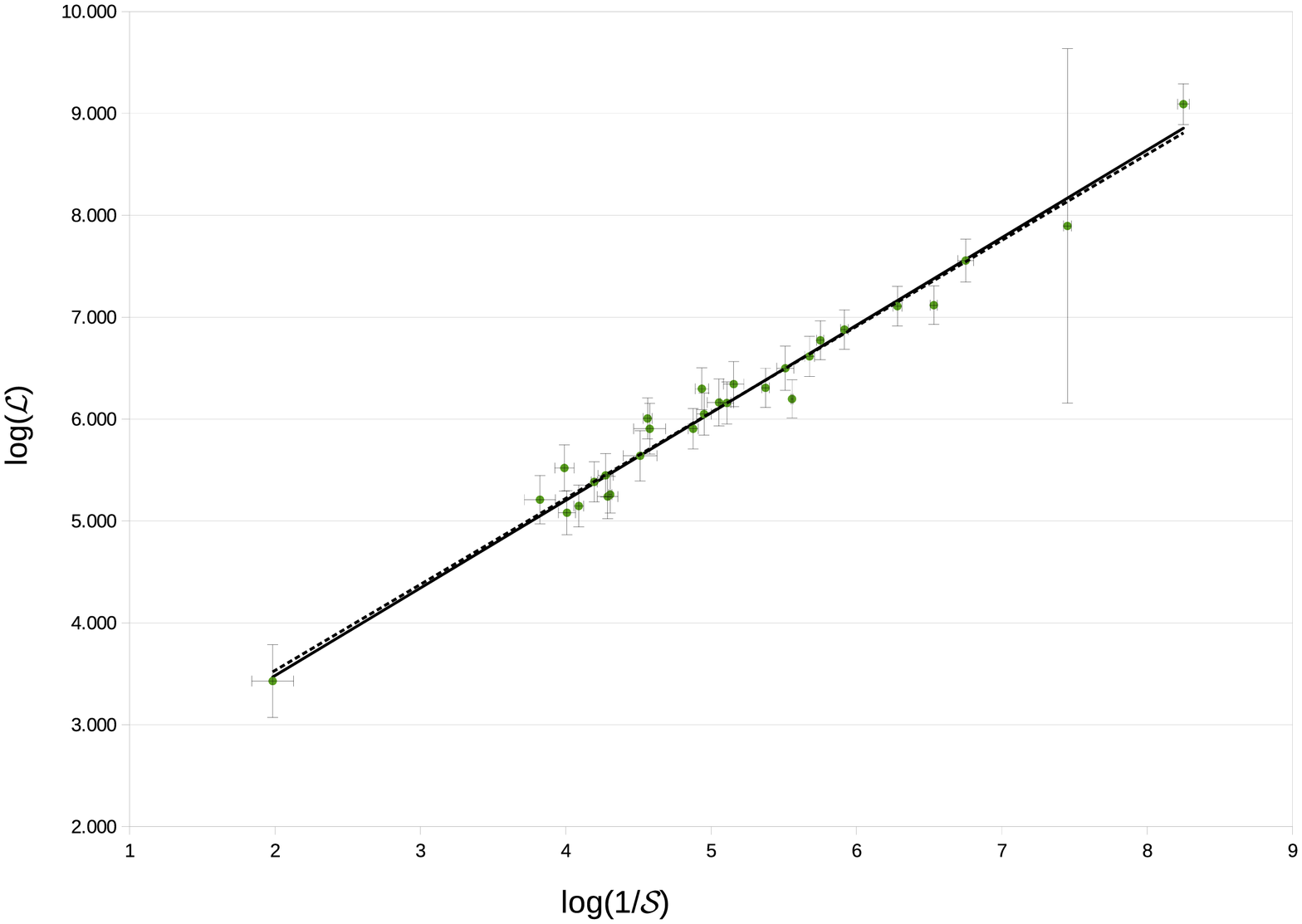}}}\\
\rotatebox{0}{\scalebox{0.50} 
{\includegraphics{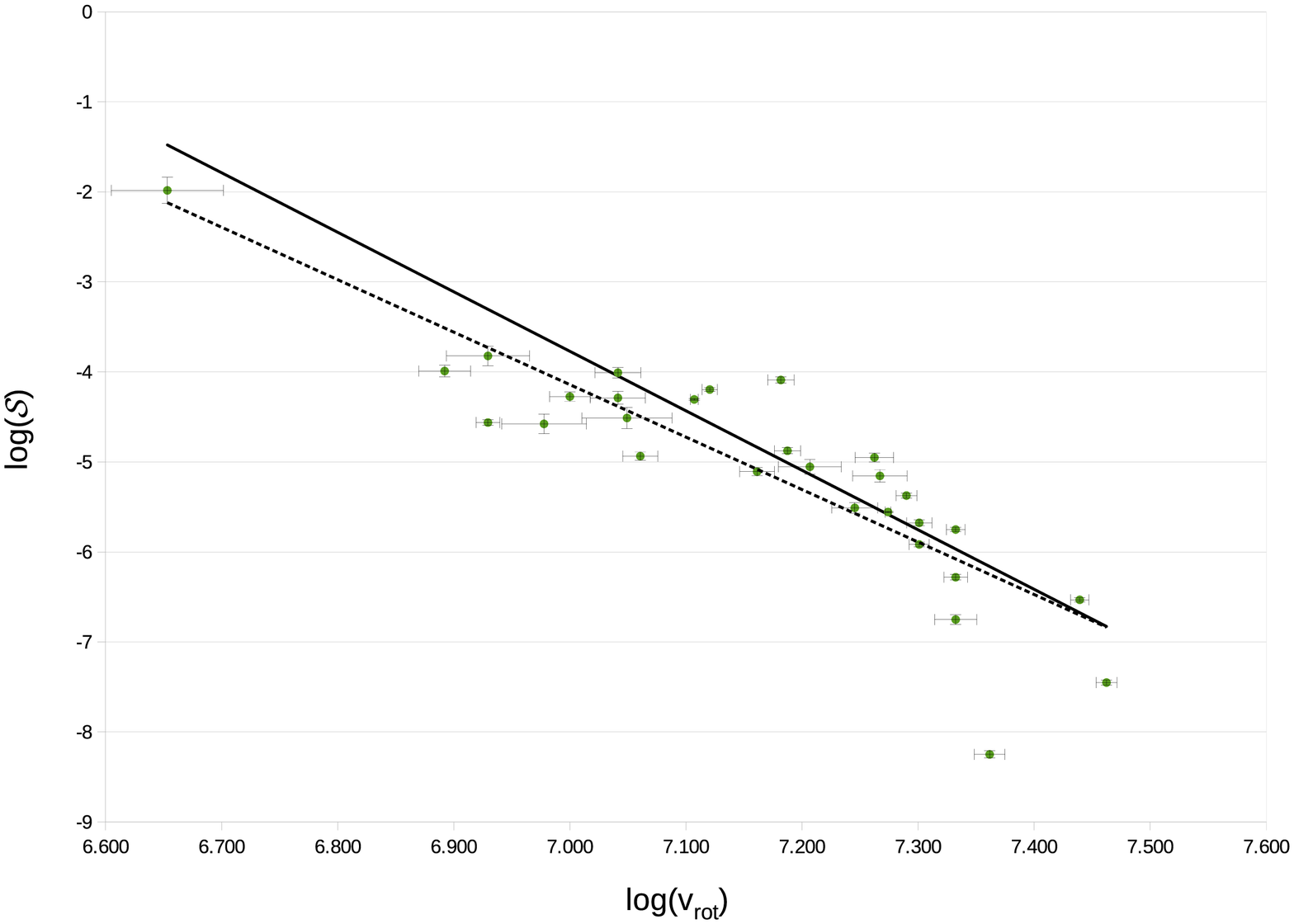}}}\\
\end{tabular}
\caption{{The top graph shows the variation of $log(\cal L)$ versus $log(1/{\cal S})$. These quantities are defined in equation \ref{eq:spiralnumbers1}.  The dashed line shows the best fit with all points weighted equally and the solid line shows the fit weighted according to the error bars of each point. The bottom graph plots $log(\cal S)$ versus $log(v_{rot})$. Dashed and solid lines are unweighted and weighted fits, respectively.}  } 
\label{fig:causality}
\end{figure}

The log plot of ${\cal L}$ versus $1/{\cal S}$ is shown in the upper panel of figure (\ref{fig:causality}).  The functional form can be considered (within errors) as an invariant property of the galaxies in the ensemble, because of the purely numerical expression. This function, using only the trend line (equal weights and dotted), is 
\be
\log{\cal L}= -0.844\log{\cal S}+1.844.\label{eq:LfctS}
\ee
The plotted errors in ${\cal S}$ and in  ${\cal L}$ are shown as they were  propagated from those given by the sample authors. The errors are slightly underestimated as we did not have error estimates for the star formation rate. The squared Pearson correlation coefficient is nevertheless $0.973$ for the unweighted trend line. 

On the same plot we show a `weighted least squares' fit to the same data \citep{JRT1997} as a solid (barely distinguishable) line. The functional form ${\cal L}=f({\cal S})$ now becomes
\be
\log{\cal L}=-0.860\log{\cal S}+1.762.\label{eq:LfctS1}
\ee
The formal error in the slope is $\pm 0.04$ and in the intercept is $\pm 0.2$. Both  versions of the function are admissible  given these errors.  Weighting by the error bars of the individual points (both in x and y as we have done) should more accurately reflect the data as presented.  An exception would be if there are systematic effects such as higher errors on one side of the plot (high or low $\cal S$ for example) compared to the other in which case the slope could be skewed when the weighting is carried out.  

This function ${\cal L}({\cal S})$ has implications for the  individual physical quantities comprising it, but it does not determine a relation between any pair of them. It takes rather the form of a reasonably well-defined numerical constraint on the variability of the three properties, $v_{rot}$, $S_*$ and $L_{3.6}$ (the luminosity is in the $3.6$ micron band). This relation is different from the bivariate relations normally sought, in that it constrains all three of the possible relations. Moreover, the slope  and intercepts are pure numbers and so have an absolute significance. 

However to apply the constraint of equation (\ref{eq:LfctS1}) we need additional information. In our discussion we take this to be the constancy of $\mu_b$, the `cosmic clock' relation \citep{Metal2018} and the lower graph  of figure (\ref{fig:causality}). In that panel we plot  $\log {\cal S}$  against $v_{rot}$ in Units of $cm~ s^{-1}$. It should be noted that this is {\it not} a true Dimensionless plot. All that we have done is to choose a Unit for $v_{rot}$ (here $cm s^{-1}$ but it might be $100~ km~ s^{-1}$), which nevertheless retains its physical Dimension of velocity.  One must distinguish physical Dimension from Unit \citep{Hen2015}

The two {plot} scales used in the graph of ${\cal S}$ against $v_{rot}$ are not identical. The errors shown for ${\cal S}$ are actually larger than those shown for $v_{rot}$. The function is not  so well defined (squared Pearson correlation coefficient $0.766$)  by the trend line as it was for the fundamental  relation (equation \ref{eq:LfctS1}), but it provides useful information.

 At present  (a larger sample is needed) the unweighted trend line takes the form (dotted line)
\be
\log{\cal S}=-5.827\log{v_{rot}}+43.65.\label{eq:Svrot}
\ee

The `weighted least squares fit' (solid line) takes the form 
\be
\log{\cal S}=-6.610\log{v_{rot}}+42.50,\label{eq:Svrot1}
\ee
where the formal error in the slope is $\pm 0.10$ and that in the intercept is $\pm 0.70$. Unlike the fundamental relation (equation \ref{eq:LfctS1}), the trend line slope lies outside these errors. The best value  of the slope of this relation may lie somewhere between the two values, perhaps close to  the average of $-6$. We suggest this in the absence of a {(preferred)} larger sample. 

 Equations (\ref{eq:LfctS}) and (\ref{eq:Svrot}) together allow  the infra red luminosity and the star formation rate to be expressed  algebraically over the sample, principally in terms of $v_{rot}$. This requires introducing Units for each physical quantity, which we take to be cgs everywhere. These become in logarithmic form  using the unweighted trend lines 
 \be
 \log{L_{3.6}}=4.09\log{v_{rot}}+14.7,\label{eq:TFir}
 \ee
 and 
 \be
 \log{S_*}=-2.83\log{v_{rot}}+43.8.\label{eq:SFR1}
 \ee
 
 Using the weighted fits in equations (\ref{eq:LfctS1}) and (\ref{eq:Svrot1}) we find these results as 
 \be
 \log{L_{3.6}}=4.07\log{v_{rot}}+14.9,\label{eq:TFir1}
 \ee
 and 
 \be
 \log{S_*}=-3.61\log{v_{rot}}+49.7.\label{eq:SFR11}
 \ee

 The first of these results is the Infra Red Tully-Fisher relation (IRTFr) that follows from our analysis. The error in the basic Number relation (\ref{eq:LfctS1}) is satisfyingly small, but the error in relation (\ref{eq:Svrot1}) is more substantial. The slope in equation (\ref{eq:SFR1}) or (\ref{eq:SFR11}) could easily  have a best value of $-3$, within the discrepancy between the trend line and the weighted fit. The average of the two values is $-3.2$. Fortunately even a slope of $-3$ makes only a rather small change in the slope of the IRTFr (equation~\ref{eq:TFir}), only increasing the slope to $4.16$. 
 
 The baryonic Tully Fisher relation (BTFr) is given by combining $\mu_b=constant$  (from figure \ref{fig:massfractions}) with the Number ${\cal V}$  and  $v_{rot}\propto R$ from \cite{Metal2018}. It  becomes (quantities in cgs Units) 
 \be
 M_b=(0.235\pm 0.023){\cal V}\frac{t_{orb}}{2\pi G} v_{rot}^3. \label{eq:BTFr}
 \ee
 From the scatter and errors given in \cite{Metal2018} the power of velocity could be as high as $3.2$.  The error in $t_{orb}$ is less than $10\%$ based on the scatter, with errors in individual samples much lower. The error in $v_{rot}$ is typically $5\%$ or less \citep{Pon2016}. Using a typical value of $2.4\times 10^{15}~s$ for $t_{orb}/(2\pi)$ gives $M_b\approx 4\times 10^9 M_\odot {\cal V}$ for $v_{rot}=10^7 $ $cm~s^{-1}$ with an error at the $30\%$ level. The value of the virial Number ${\cal V}$ is expected to be $O(1)$, depending on the radius chosen \citep{BT2008}.  
 
  This is similar to the result given by the authors of this sample. Bigger but equally accurate samples are required, but provisionally we conclude that {\it the  BTFr may have a distinctly different origin from that of the IRTFr}. As derived here the BTFr, appears to arise from the dynamical structure at the edge of the galaxy combined with constant baryonic mass fraction.
  
\begin{figure} 
\begin{tabular}{cc} 
\rotatebox{0}{\scalebox{0.50} 
{\includegraphics{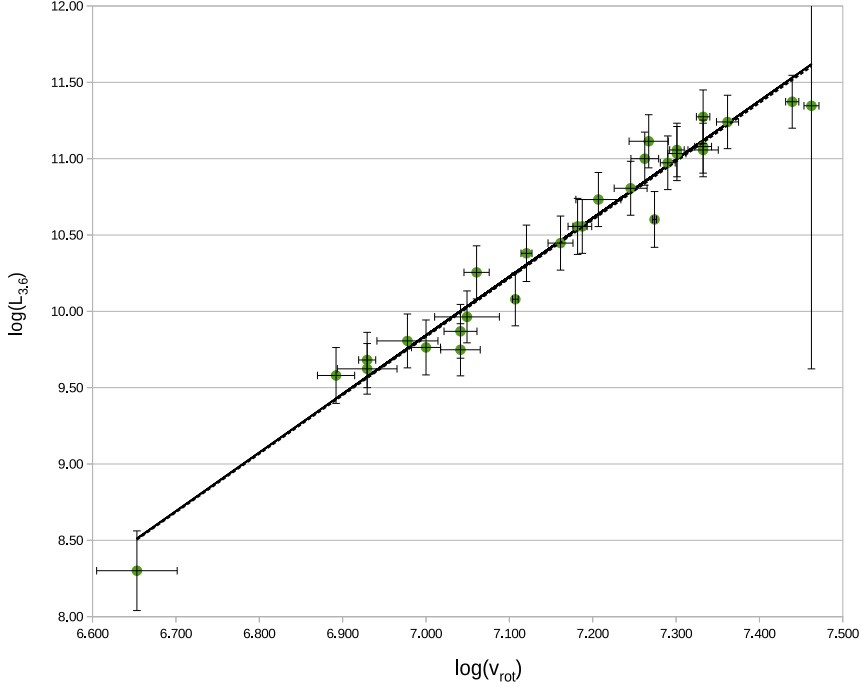}}}
\end{tabular}
\caption{We show for the same sample the standard Tully-Fisher plot using the infra red ($3.6 \mu$) luminosity.  } 
\label{fig:StdTF}
\end{figure}

  A comparison of  method advocated above with the standard Tully-Fisher approach is desirable. We accomplish this by plotting the luminosity versus the rotation velocity on a log-log plot. This is shown in figure (\ref{fig:StdTF}). The weighted fit to the data is 
  \be
  \log{L_{3.6}}=(3.839\pm0.229)v_{rot}-(17.030\pm 1.664),\label{eq:StdTF1}
  \ee
  while the trend line yields essentially the same fit with the  squared Pearson correlation coefficient equal to $0.962$. The slope is equal to that found above within errors, and both slopes could be equal to $4$. The scatter is admirably small. 
  
  The errors here are slightly better controlled than the errors in either equation (\ref{eq:TFir}) or (\ref{eq:TFir1}), but we note that they are  essentially comparable to those in the `guiding' relation (\ref{eq:LfctS1}). Surprisingly, they are rather larger than the statistical errors associated with relation (\ref{eq:Svrot1}). 
  
  It is clear that the main benefit of our approach is organizational. The fundamental Number relation (\ref{eq:LfctS1}) is a numerical  absolute (the intercept in equation (\ref{eq:StdTF1}) is Units dependent) that indicates what additional information is required. We have  also inferred the constancy of the Number $\mu_b$ and consequently the form of the BTFr. Moreover equation (\ref{eq:SFR11}) suggests a slightly unexpected relation.

  \newpage

 Relation (\ref{eq:SFR11}) gives an inverse dependence of the global star formation rate (SFR) on the outer flat rotation velocity. This may seem surprising at first glance, but replacing $v_{rot}$ by $R$ according to \cite{Metal2018} shows that the dependence is really on inverse radius. It is then a dependence on `concentration' , assuming roughly constant density $M_b/R^3$.  The nominal value $-2.83$  or $-3.61$ of the slopes could  be close to $-3$ within errors, in which case the SFR is inversely proportional to the volume of the galaxy. 
 
 It is interesting to note that one can obtain another estimate for the star formation rate by combining the Numbers ${\cal S}$ and our neglected $Q$. The Number ${\cal F}\equiv {\cal S}/Q^3 \equiv GS_*/\sigma^3$. This Number is significant because the value of ${\cal F}$ is increased by roughly three orders of magnitude and its variation is reduced dramatically. By the filters we used above this suggests that ${\cal F}\approx constant$ over the ensemble, although we do not have the data to prove this result. Hence we speculate that 
 \be
 S_*={\cal F}\times \frac{\sigma^3}{G}.\label{eq:SFR2}
 \ee
We can not check this estimate on the sample since $\sigma$ is not given for each galaxy. However with $\sigma=3.2\times 10^6 ~cm ~s^{-1}$ \citep{Pon2016}, ${\cal F}=O(2\times 10^{-3})$ (assuming $v_{rot}=1.6\times 10^7~ cm~ s^{-1}$) we obtain a number in cgs Units comparable to that of equation (\ref{eq:SFR1}). Equating the two expressions formally, suggests that $\sigma$ and $v_{rot}$ are nearly inversely correlated over the sample.     
 
The most satisfying result of this section is the constraint (equation \ref{eq:LfctS1}) as indicated graphically on the {top} in figure (\ref{fig:causality}). The graph on the {bottom} in the same figure is subject to more error, which arises mainly from the error in $v_{rot}$.  
The scatter is greatly reduced in the Numbers plot, presumably because there are fewer missing variables.  In our concluding section we summarize our conclusions and give some suggestions as to the physical meaning of the various Numbers.

\section{Summary Discussion}
 
 In general when adopting a positivist or measured description of a spiral galaxy, the catalogue might look something like
 \be
 SC:=\{L_\nu,S_*,v_{rot}, \sigma, M_b, M_*, M, R,B,G,z\},\label{eq:genSC}
 \ee
 where we have placed the less directly observed quantities towards the end together with the more platonic aspects ($G,z$). Such a catalogue is what we have referred to as `hybrid', and it would be even larger if synchrotron radiation for example were generally observed.  In specific observational sets the galaxy description is normally less complete than this, and we are forced to deal with the `observer's galaxy' and create an effective hybrid catalogue. The catalogue of the previous section lacked a measure of velocity dispersion for example. 
 
 Proceeding for the moment to construct the Numbers following from equation (\ref{eq:genSC}), we should nevertheless remember that they are not unique, although they are complete given the catalogue. In particular Numbers can be constructed sometimes from the complete set, so that they are large, small or $O(1)$, when they may not have been in the first choice of Numbers. This allows for simplifications as we have seen. However following the pattern of our main text, the catalogue (\ref{eq:genSC}) implies a total of eight Numbers, of which the red shift $z$ can be one. We choose them to be the set
 \bea
 {\cal L}&\equiv& \frac{L_\nu}{S_*v_{rot}^2},~~~~{\cal S}\equiv \frac{GS_*}{v_{rot}^3},~~~~{\cal F}\equiv \frac{GS_*}{\sigma^3},~~~~z,\nonumber\\
 {\cal M}&\equiv& \frac{R^3B^2}{Mv_{rot}^2},~~~~{\cal V}\equiv \frac{GM}{Rv_{rot}^2},~~~~\mu_*\equiv \frac{M_*}{M},~~~~\mu_b\equiv \frac{M_b}{M}.\label{eq:genNumbers}
 \eea
We note that an example of a constructed quantity is the Toomre Number $Q=\sigma/v_{rot}$, or some power thereof. Clearly $Q^3={\cal S}/{\cal F}$, and $Q$ could be used in place of either ${\cal S}$ or ${\cal F}$, but we continue as above to follow the pattern of our galaxy example. 

We have argued in the previous section that the Numbers $\mu_*,\mu_b,{\cal M}, {\cal V},{\cal F}$ have variations of $O(1)$, to which may be added {the redshift}, $z$  (or a function of $z$), for nearby galaxies. This left us with the well defined constraint ${\cal L}=f({\cal S})$ in equation (\ref{eq:LfctS1}), which we recall here as 
\be
\frac{\cal L_\nu}{S_*v_{rot}^2}=C_1\big(\frac{GS_*}{v_{rot}^3}\big)^{-0.860},\label{eq:LfctS2}
\ee
where the constant $C_1$ is known. 

However the more traditional relation (\ref{eq:SFR11}) gave us approximately 
\be
{\cal S_*}=C_2~10^{49.7}\big(v_{rot}\big)^{-3.610},\label{eq:S*rot2}
\ee
where $C_2$ is also known and all quantities are cgs Units.  Together these last two equations give  our principal result for the IRTFr
\be
L_\nu=10^{6.36}\frac{C_1C_2^{0.14}}{G^{0.860}}~v_{rot}^{4.08}.\label{eq:TFir2}
\ee
If the power in equation (\ref{eq:S*rot2}) is $-3$, then $L_\nu\propto v_{rot}^{4.16}$. 

We pointed out that constant $\mu_b$ plus $R\propto v_{rot}$ over the sample \citep[e.g.][]{Metal2018} suffices to give the BTFr 
\be
M_b\propto v_{rot}^3,\label{eq:BTFr}
\ee
which suggests a possible cause independent of the traditional Tully Fisher relation. 

There are various ways to interpret the Numbers as ratios of physical quantities. Indeed this is the  way of tracking the physical explanations for the various correlations. We suggest here some possibilities although they are not unique but are possibly complete. The Numbers ${\cal V}, \mu_*, \mu_b, {\cal M}, z$ are self-evident. We have already interpreted the Toomre Number $Q$ in equation (\ref{eq:Toomre}). If we write $S_*=\Delta M_*/\Delta t$ then 
\be
{\cal F}=\frac{G\Delta M_*}{\sigma^2}\frac{1}{\sigma\Delta t}.\label{eq:physF}
\ee
This appears as the ratio of the gravitational radius of the new stars in a medium of turbulent dispersion $\sigma$ divided by the distance covered by a turbulent element in the formation time $\Delta t$. One can see intuitively that if this Number is too small, star formation will be limited.  It appears to be $O(1)$ in the sample of the previous section. 

The Number ${\cal S}=Q^3{\cal F}$, which we can expect to be small in the presence of star formation because $Q^3\ll 1$, which accords with the condition for gravitational instability (\ref{eq:Toomre}). By analogy with ${\cal F}$, ${\cal S}$ can also be seen as the gravitational radius of the new stars in a medium with velocity $v_{rot}$ divided by the distance travelled in the formation time at speed $v_{rot}$. This is certainly small as we have found. If we use the Number ${\cal V}$ to eliminate $v_{rot}$ one can find that ${\cal S}=\Delta\mu_*/(\Delta t \sqrt{G\rho}){\cal V}^{3/2}$, where $\rho=M/R^3$. This is the star formation rate as a fraction of the dynamical mass, multiplied by the  the mean gravitational time (free fall or rotation period) of the galaxy. Simple estimates show again that it is likely to be very small and to vary widely.

Finally the Number ${\cal L}$ can be written by eliminating $v_{rot}$ as 
\be
{\cal L}=\frac{L_\nu}{(G\Delta M_*M/R\Delta t)}{\cal V},\label{eq:physL}
\ee
which is the ratio of galactic energy radiated per unit time in the infra red band to the  gravitational energy  associated with  the new stars created in the stellar formation time. Dividing  the original form of ${\cal L}$ by $Q^2$ allows us to see the resulting Number as the ratio of the luminosity in the infra red band  to the turbulent energy tied up in  the motion of new stars in a  stellar formation time. Both versions of this Number are expected to be large and to vary widely.

\section{Note Added}

A referee has pointed us to a paper remarkably similar to this one, written by G.S. Golitsyn in 2013 \cite{G2013}. We were not aware of this paper when we were writing the current version of the same general idea. It should perhaps be regarded as some support for the procedure given two completely indepedent origins.

\label{lastpage}
\end{document}